\documentclass[showpacs]{revtex4-1}
\usepackage{amsmath,amssymb}
\usepackage{bm}
\usepackage{ulem}
\usepackage[]{graphicx}

\usepackage{color}

\begin{document}

\title{Dynamic oscillatory cluster ordering of self-propelled droplets}

\author{Shinpei Tanaka}\email{shinpei@hiroshima-u.ac.jp}
\affiliation{Graduate School of Integrated Arts and Sciences, Hiroshima
University, 1-7-1 Kagamiyama, Higashi-Hiroshima 739-8521, Japan}

\author{Takeshi Kano}
\affiliation{Research Institute of Electrical Communication, Tohoku University,
2-1-1 Katahira, Aoba-ku, Sendai  980-8577, Japan}

\pacs{
      68.03.Cd 	
      82.40.Bj 	
      82.40.Ck 	
      05.65.+b 
}

\begin{abstract}
 We report here a peculiar dynamically ordered state of clustering
 droplets of a mixture of organic solvent. There droplets are driven by
 the solutal Marangoni effect on the surface of aqueous surfactant
 solution. They form temporal ring clusters which start collapsing
 immediately after its formation. This process is repeated for more than
 several hours with the period of 5--20 minutes. We propose an
 inhomogeneous force model to phenomenologically understand the basic
 mechanism of this dynamics, where the forces acting on each particle
 are controlled differently. This droplet system offers a simple,
 non-biological experimental model for the study of complex dynamical
 states realized by a group of self-propelled particles.
\end{abstract}
\maketitle

Biological functions sometimes arise from a cluster of
active elements. For example, it has been proposed that the swarming
behaviors of bacteria is beneficial to each bacterium in an occasion
such as a food intake or resisting starvation \cite{Kaiser2003}. There
the motility and communication of bacteria play important roles to
initiate coordinated behaviors. It has been proposed that the complex
interaction among them has been evolving to optimize the required
functions, which may have even led to the multicellularity
\cite{Shapiro1998}.

To fully understand these collective behaviors seen in biological
systems, it is important to construct non-biological systems capable to
model them. It is now well-known that the swarming phenomena are not
limited in biological systems, but can be also
seen in many non-biological motile systems
\cite{Sumino2012,Thutupalli2011,Bechinger2016}. Their connection to the
biological systems has been studied by many researchers so far
\cite{Marchetti2013}.

Beyond the relatively simple pattern formations observed so far in the
non-biological systems, however, it is a challenging task to find
physical systems capable to exhibit complex and dynamical collective
patterns comparable to biological systems. Ikura et al. found the
collective motion of camphor boats, exhibiting a billiard-like transfer
of motion if confined in a narrow channel
\cite{Ikura2013,Nakata2015}. Soh et al. reported that gel particles
containing camphor created characteristic dynamical ordering
\cite{Soh2011}.

We have recently shown that droplets of alkyl salicylate floating on
aqueous surfactant solution exhibit a stable self-propulsion
\cite{Tanaka2015}. Here we report collective behaviors of similar
droplet systems, where droplets show much more complex modes of motion
than those observed in non-biological systems reported so far. Our
droplets contain two liquids, ethyl salicylate (ES) and paraffin liquid,
and the dissolution of ES produces a surface tension gradient around the
droplet that propels it \cite{Tanaka2015}. Thus ES is used as a fuel,
whereas inert paraffin liquid composes a droplet body. Although this droplet
system consists of simple components, it can exhibit various complex
collective behaviors, which actually resemble wiggling motion of
earthworms.

In this report, we propose {\em inhomogeneous} forces as a tool to
understand the complex behaviors of droplets. Our droplets interact with
each other via the field around them created by themselves. There the
field created by an isolated particle is not necessarily the same with
the one created by the same particle in a cloud of particles. In other
words, the superposition principle of the interaction does not
necessarily hold. Therefore this way of the interparticle interaction
can be beyond {\em homogeneous}; homogeneous here means that all the
elements share the same type of interaction depending only on the
distance between them. The interaction becomes {\em inhomogeneous} if
the local environment around an element differs from others. In the
first part of this report, we describe the experimental observations of
our droplet system. Then in the second part, we show that a simple
inhomogeneous model can reproduce the observations at least in part.

Droplets of ethyl salicylate (ES, Tokyo Chemical Industry, Tokyo) used
in this study contained 30 wt\% of paraffin liquid (Sigma-Aldrich,
Tokyo). The volume of a droplet was 10 $\mu$l. The droplets were placed
on the surface of 30 ml aqueous sodium dodecyl sulfate (SDS, Tokyo
Chemical Industry) solution in a glass dish of 86 mm in inner
diameter. The concentration of SDS was fixed at 35 mM. A glass cover was
used without a tight sealing. The number of droplets, $N$, was
$1-200$. The elapsed time, $t$, was measured from the time when all the
droplets were placed on the aqueous surface. The droplets were dyed with
Oil red O (Nakarai tesque, Tokyo). The chemicals were used as supplied.

The motion of droplets was recorded using a CMOS camera (L-835, Hozan,
Osaka) as a movie. The sample as well as the camera were placed in a
temperature-controlled box, where the temperature was controlled at
$21.0\pm 0.2^\circ$C using a heater. From a recorded movie, images of
640$\times$480 pixels were extracted and the position of droplets was
detected using a software, ImageJ (http://rsb.info.nih.gov/ij/). By
differentiating the position numerically, the velocity was calculated.

 \begin{figure}
  \begin{center}
   \includegraphics[width=14cm]{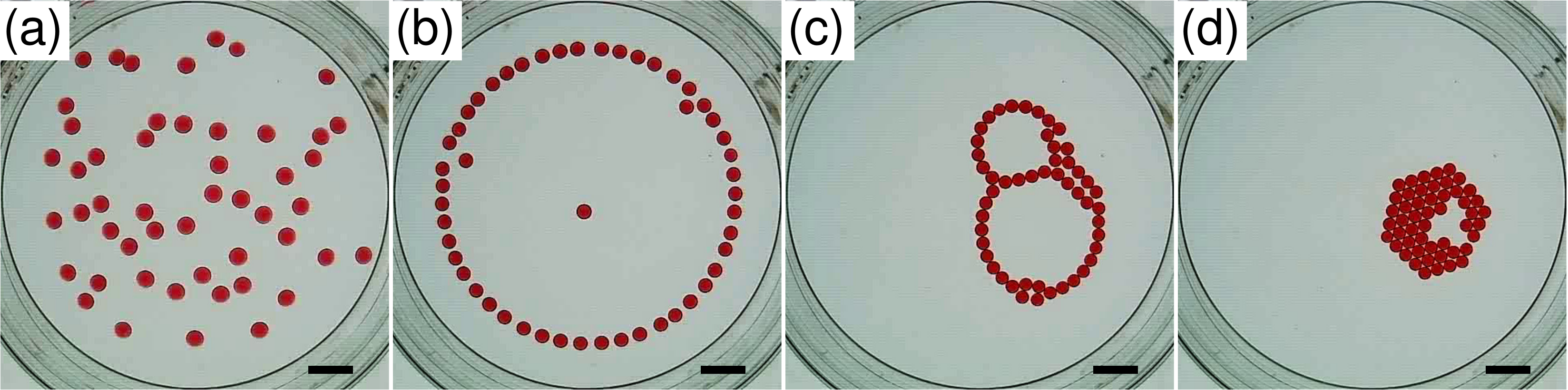}
   \caption{The modes of motion seen in a system of $N=50$ in
   sequence. (a) Random, intermittent motion. (b) A ring of droplets where
   droplets vibrated between others. (c) The characteristic cluster
   oscillation. (d) A final static crystal. The scale bar is 10 mm.}
   \label{process}   
  \end{center}
 \end{figure}

Figure \ref{process} shows four typical structures seen in sequence in a
system of $N=50$ (see also the Supplementary Material, SM). At first the
droplets are in random, intermittent motion [Fig.~\ref{process}(a)
and mov-1, SM] but gradually form a ring where each droplet
vibrates between the others [Fig.~\ref{process}(b) and mov-2,
SM]. We refer in this report that this oscillation of a single droplet
as droplet oscillation. The period of droplet oscillation is typically
several seconds. The ring shown in
Fig~\ref{process}(b) is the result of a circular dish used. When in a
larger square dish, the droplets form clusters of typically $N\sim 10$
(mov-3, SM).

Then the ring collapses into a cluster and a periodic wiggling motion of
the cluster begins [Fig.~\ref{process}(c) and mov-4, SM]. We refer this
oscillatory motion of a cluster as cluster oscillation. This cluster
oscillation can be observed in a cluster as small as $N=5$. When $N$ is
larger than about 10, this cluster oscillation looks a periodic wiggling
motion. The period of cluster oscillation is about 10 minutes, much
longer than that of droplet oscillation.  This cluster oscillation was
also observed in a smaller (60 mm) and larger (200 mm) dish (mov-5,
SM). Finally, the droplets were aligned on a hexagonal lattice to form a
crystal with well-defined facets [Fig.~\ref{process}(d) and mov-6, SM].

 \begin{figure}
  \begin{center}
 \includegraphics[width=14cm]{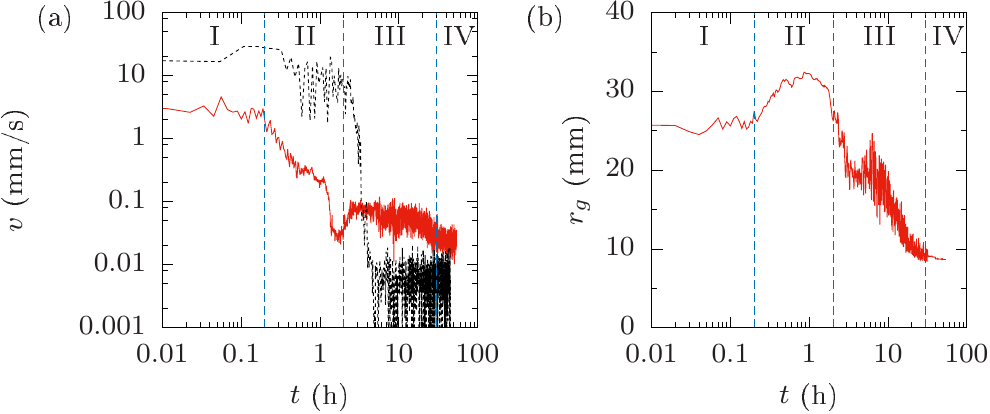}
 \caption{(a) The average speed of droplets, $v$. The solid red line
 shows the values when $N=50$; the dashed black line shows the values
 when $N=1$ under the same condition. (b) The radius of gyration, $r_g$,
 of the droplet distribution. The stages I to IV corresponds with the
 motion patterns shown in Fig.~\ref{process}.}  \label{absv}   
  \end{center}
 \end{figure}

\begin{figure}
 \begin{center}
 \includegraphics[width=14cm]{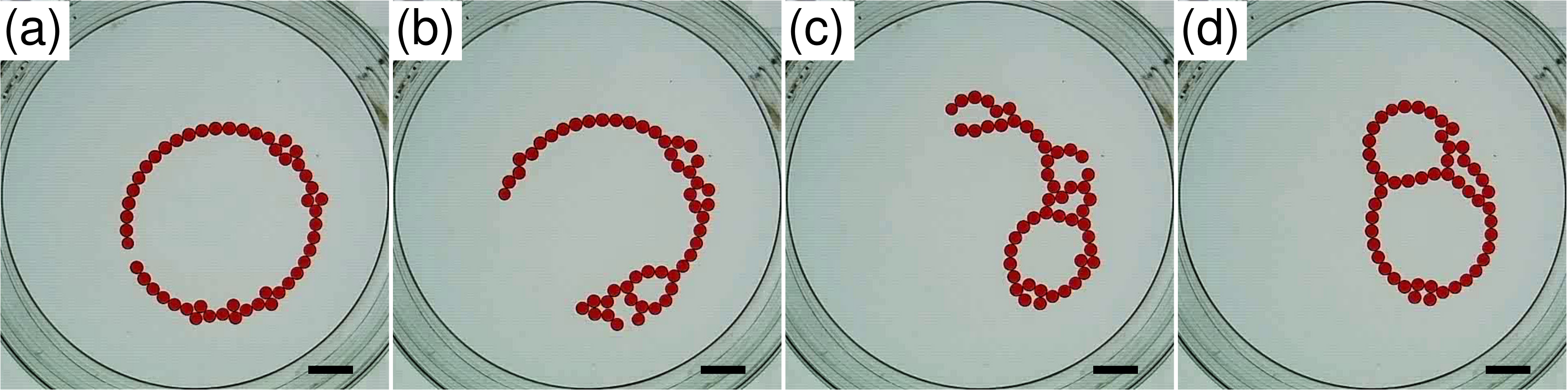} \caption{Snapshots of
 droplets in the cluster oscillation ($N=50$). The elapsed time was
 $t=8$ h. The time elapsed from (a) was, (a) 0 s, (b) 116 s, (c) 336 s,
 and (d) 504 s. The scale bar is 10 mm.} \label{worm}  
 \end{center}
\end{figure}

Figure \ref{absv}(a) shows the average speed of a droplet in a system of
$N=50$ (red solid line) and of an isolated droplet ($N=1$, black dashed
line) under the same condition. It shows that the droplets' motion was
suppressed at first when many droplets coexist. However, even after
the isolated droplet stopped, droplets in a system of $N=50$ kept
moving, though slowly. This slow movement corresponds to the cluster
oscillation shown in Fig.~\ref{process}(c) and Fig.~\ref{worm}.

Figure \ref{absv}(b) shows the radius of gyration, $r_g$,
calculated as the root mean square distance of all droplets
from the center of mass of the droplet distribution. It showed a
maximum and then decreased to the value of a compact cluster shown in
Fig.~\ref{process}(d). Using Fig.~\ref{absv}(a) and (b), we divide the
process of dynamic ordering into four stages, where stages I to IV
correspond with the patterns (a)-(d) shown in Fig.~\ref{process}. The
reproducibility of the process was high though the time of transition
from a stage to the next depends on the conditions. In the stage I, $v$
is the highest and the droplets are randomly distributed. In the stage
II, $v$ decreases rapidly and $r_g$ reaches the maximum reflecting the
ring structure along the wall [Fig.~\ref{process}(b)]. After $v$ reaches
the minimum, the ring collapses and the cluster oscillation begins
[Fig.~\ref{process}(c)] in the stage III. There $v$ is recovered to a
certain value. The stage III continued for about 30 hours in the sample
shown in Fig.~\ref{process}. During the stage III, $r_g$ keeps
oscillating. In the late stage of the stage III, the wiggling cluster
shrinks gradually. In the stage IV, the cluster becomes a crystal.

Figure \ref{worm} shows snapshots of the cluster oscillation seen in a
system of $N=50$ (mov-4, SM). When $N\gtrsim 10$, the motion pattern
is in general the repeated formation and collapse of rings as
follows. First, a ring cluster (which was smaller than the one in the
stage II) is formed with a thread of droplets [Fig.~\ref{worm}(a)]. It
breaks at a point [Fig.~\ref{worm}(a)], then the thread starts shrinking
with forming local small rings within it [Fig.~\ref{worm}(b)]. The local
rings sometimes merge while another rings are forming continuously
[Fig.~\ref{worm}(c)]. Finally one ring becomes dominant and a
large ring cluster is recovered [Fig.~\ref{worm}(d)]. Then the
process is repeated.

\begin{figure}
 \begin{center}
 \includegraphics[width=14cm]{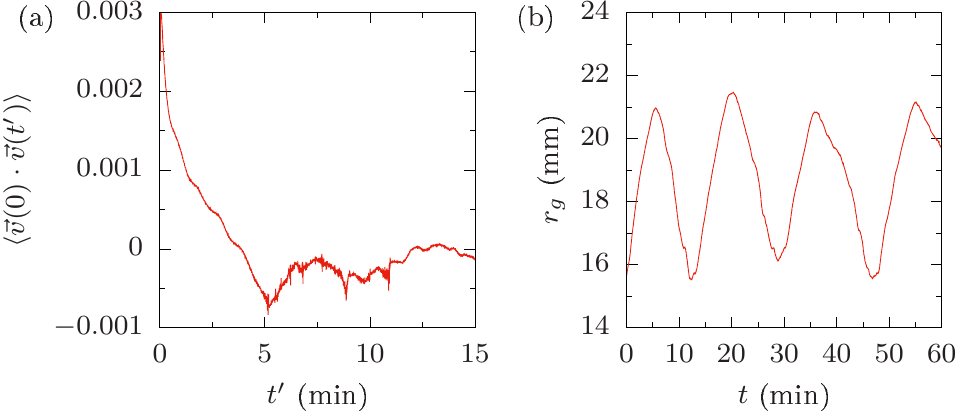} \caption{(a)
 The velocity autocorrelation function of a droplet in the cluster shown in
 Fig.~\ref{worm}. (b) The oscillation of $r_g$ of the same
 cluster. $t=0$ (min) here corresponds with $t=8$ h in Fig.~\ref{absv}.}
 \label{oscillation}  
 \end{center}
\end{figure}

Figure~\ref{oscillation} shows the oscillation of the cluster shown in
Fig.~\ref{worm}. The velocity autocorrelation function
[Fig.~\ref{oscillation}(a)] has a negative part after a rapid decrease
at around $t'\simeq 5$ (min), reflecting the circular motion of a droplet
in the cluster. The radius of gyration of the cluster
[Fig.~\ref{oscillation}(b)] exhibits clear oscillation, with the period
of about 15 minutes in this case. 

\begin{figure}
 \begin{center}
 \includegraphics[width=14cm]{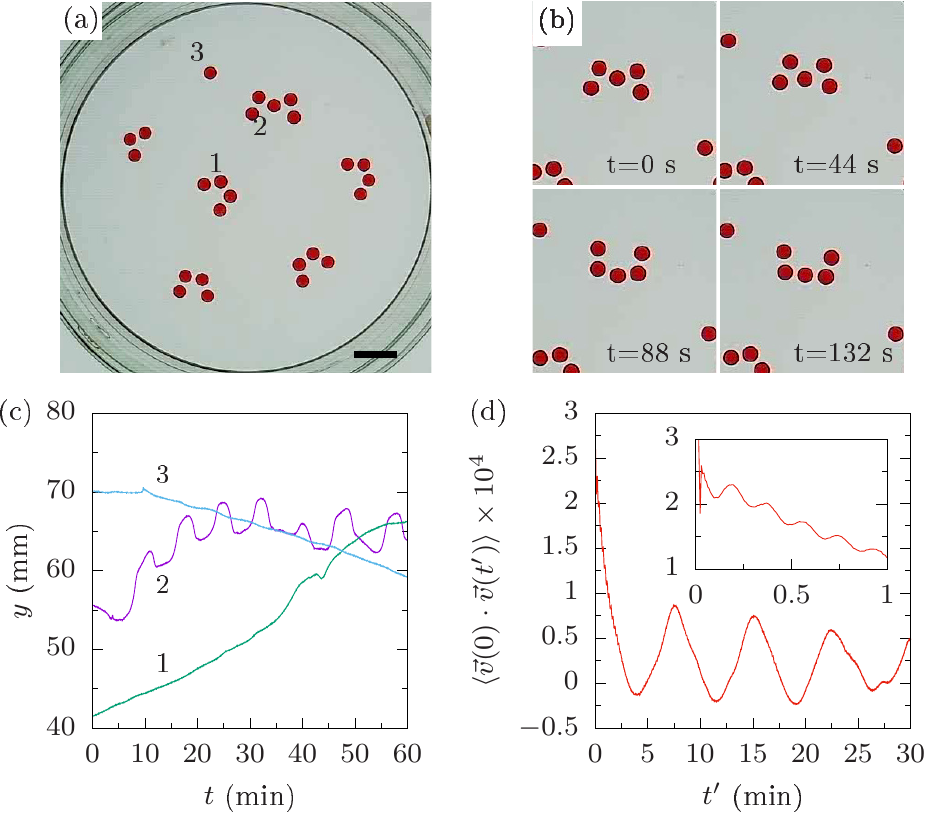}
 \caption{The dynamic states of small clusters ($N=25$). (a) Clusters
 observed at $t=5$ h. The scale bar is 10 mm. (b) The cluster
 oscillation in a cluster of $N=5$. (c) The change in the $y$-coordinate
 of the droplets. The figures represent a droplet shown in (a). (d) The
 velocity autocorrelation function averaged over all the droplets. Inset
 is an enlarged view in the region of $t'<1$ min.}  \label{small}  
 \end{center}
\end{figure}

The complex cluster oscillation (periodic wiggling) may be decomposed
into simpler ones, if $N$ is small. Figure \ref{small}(a) shows the
small clusters formed in a system of $N=25$. Other conditions were the
same as the one shown in Fig.~\ref{process}. The motion of the cluster
``2'' is depicted in sequence in Fig.~\ref{small}(b), where a droplet in
between four droplets moved back-and-forth with the period of about 10
minutes [Fig.~\ref{small}(c), ``2'' and mov-7, SM]. Although it is
not like ``wiggling'' for this small cluster, it is a minimum cluster
oscillation with the similar period of oscillation.

The oscillation of this small cluster was also seen clearly in the
velocity autocorrelation function [Fig.~\ref{small}(d)]. Moreover, it
shows another rapid oscillation [Fig.~\ref{small}(d), inset] with the
period of about 10 s. This faster oscillation corresponds with a
remaining droplet oscillation seen mainly in the stage I and II, which
decays gradually while the cluster oscillation persists.

On the other hand, the cluster with four droplets, assigned with ``1'',
was in a translational motion [Fig.~\ref{small}(c), ``1'']. An isolated
droplet, ``3'', was in a slower translational motion. Thus the
clusters can be in a different dynamical state even if they are in the
same dish.

Let us summarize the observation on the cluster oscillation and add some
additional information. (1) The cluster oscillation is a slow mode of
motion that appears when the unbalanced surface tension gradient, that
is, the Marangoni effect, mostly decays. Only the existence of
neighboring droplets can induce the net motion. (2) The cluster
oscillation is long-lasting motion. In the example shown in
Fig.~\ref{worm}, it continued about 30 hours; in the example shown in
Fig.~\ref{small} it lasted for more than two weeks. (3) The cluster
oscillation is robust against $N$, when the droplets can interact with
each other. We observed this mode of motion when $N=5-200$ so far.

The observations (1) to (3) suggest that the cluster oscillation is
caused by the characteristic interaction among droplets after the main
self-propulsion force decays. Judging from the complex cluster motion
observed, however, this interaction is not the simple one depending only
on the relative arrangement between two particles, but is active one
depending also on the local environment and possibly on the internal
state of the particle. This is crucially different from solid particle
systems such as camphor boats, where it is difficult for the elements to
have the time-dependent internal state. Moreover, it has been proposed
that the existence of internal states can produce variety of dynamic
patterns \cite{Tanaka2007}.

Regarding the interaction, the experiments suggest that there are at
least three types of interaction among droplets: (a) an attraction
between droplets due to the capillary interaction, which eventually
stabilizes close-packed, crystalline aggregates; (b) a short-range
(possibly microscopic) unknown mechanism preventing the droplets'
coalescence; (c) a time-dependent interaction causing the cluster
oscillation.

There is additional information regarding (c). When the droplets are
confined in a linear channel, they do not form a chain but keep a
certain distance with occasional droplet oscillation (mov-8 and mov-9,
SM). Since the Marangoni flow is restricted to the direction parallel to
the channel and is known to induce repulsion between particles
\cite{Soh2011}, it is the confinement that makes the
Marangoni-flow-induced repulsion dominant, and prevent the cluster
formation. Therefore this suggests that the interaction (c) consists
partly of the directional Marangoni-flow-induced repulsion.

It is considered that the interaction (c)
reflects the surrounding field produced as a consequence of the history
of droplets' motion. Thus it seems not possible to reproduce the
observed phenomena by assuming that all the particles interact with each
other via homogeneous interaction rules. Therefore, we test a
simple phenomenological model where each elements interact with each
other differently. This approach will tell us important clues to
understand the essential mechanisms of the cluster oscillation.

Under the assumption of the overdamped limit, we model the equation of
motion of a particle as in dimensionless form,
 \begin{align}
  \label{eom r}
  \bm{v}_i&=\sum_{j\neq
 i}(-k_{ij}R_{ij}^{-1}+R_{ij}^{-2})\bm{e}_{ij}+f_i\bm{n}_i\\
 \label{eom n}
 \tau\dot{\bm{n}}_i&=\left\{\bm{n}_i \times (\bm{v}_i/v_i)\right\}\times \bm{n}_i,
 \end{align}
where $\bm{v}_i$ is the velocity of $i$-th particle and $\bm{e}_{ij}$ is
the unit vector pointing from $j$-th particle to $i$-th
particle. $R_{ij}$ and $k_{ij}$ are the distance and the attraction
strength between $i$-th particle and $j$-th particle, respectively. We
assume simple power law attraction and repulsion between particles. The
propulsion force, $f_i$, acts in the direction of the unit vector
$\bm{n}_i$ that represents the internal state of $i$-th particle. We
adopt a model proposed by Shimoyama et al. \cite{Shimoyama1996} for the
change of $\bm{n}_i$ [Eq.~\eqref{eom n}] with the relaxation time
$\tau$.

The inhomogeneity enters in $k_{ij}$ and $f_i$,
thus the number of parameters increases
quickly with $N$. The model, therefore, is applicable only to small
clusters, and is useful to see how the inhomogeneous forces induce their
cluster oscillation.

We nondimensionalized the variables using the diameter of a droplet
($d^\ast=3\times 10^{-3}$ m), the average speed ($v^\ast=2\times
10^{-5}$ m/s), and the friction constant ($\zeta^\ast=1\times 10^{-5}$
kg/s) estimated according to the Stokes law. Then the unit of time was
$t^\ast=150$ s and the typical force, $f^\ast=\zeta^\ast v^\ast$, was
nondimensionalized as 1. The equations \eqref{eom r} and
\eqref{eom n} were solved using the fourth order Runge-Kutta method with
the time step of 0.001.

 \begin{figure}[htbp]
  \begin{center}
 \includegraphics[width=14cm]{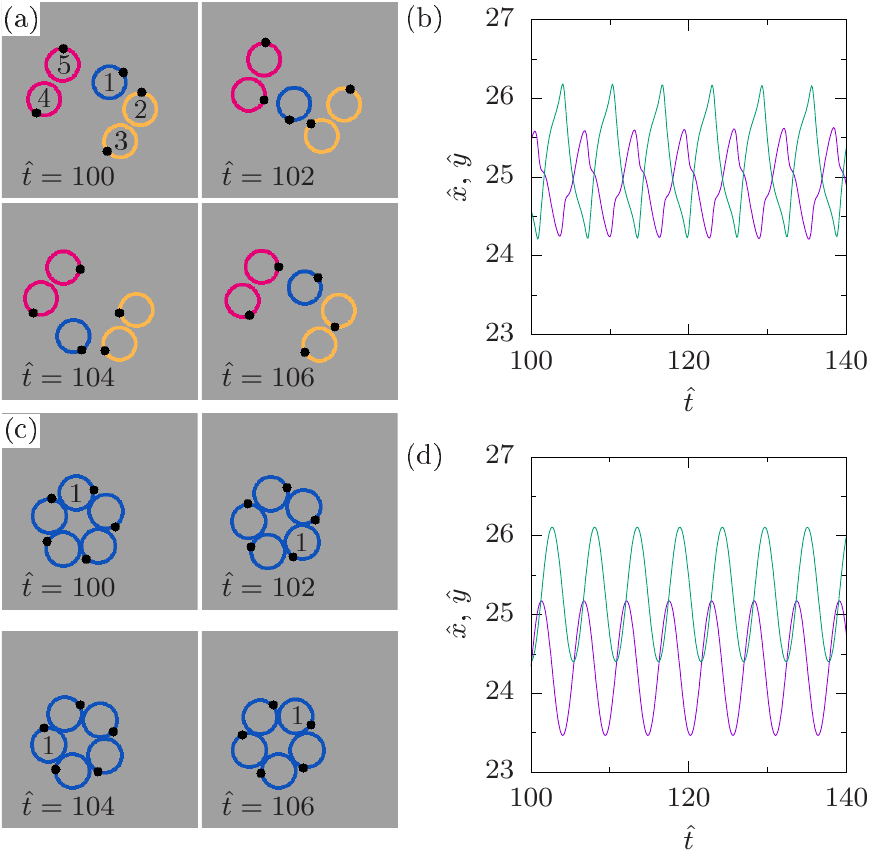}
 \caption{A 5-element cluster obtained by the simulation with the
 inhomogeneous forces, (a), and with the homogeneous forces,
 (c). $\tau=0.1$. The inhomogeneous interaction in (a) is: $k_{ij}=0.8$
 for $i=1$ or $j=1$, or $(i,j)=(2,3),(3,2)$, or $(i,j)=(4,5),(5,4)$;
 $k_{ij}=0.3$ for other combinations of $(i,j)$. The propulsion force
 was $f_1=1.0$ and $f_i=0$ for $i\neq 1$. In (c), $k_{ij}=0.8$ for
 $i\neq j$, and $f_i=1.0$ for all $i$. The $\hat{x}$ and $\hat{y}$
 coordinates of the blue element ($i=1$) in (a) are plotted in (b), and
 those in (c) are plotted in (d). A
 black dot on the elements represents $\bm{n}_i$. } \label{simulation}   
  \end{center}
 \end{figure}

Figure \ref{simulation} shows an example of a 5-element system. In
Fig.~\ref{simulation}(a), the attraction between pink particles and
orange particles was weaker than others, and only the blue particle was
propelled. Then the particles self-organized into an oscillatory state
similar to the one shown in Fig.~\ref{small}(b)
[Fig.~\ref{simulation}(b) and mov-10,
SM]. This oscillatory state was not sensitive
to the difference of attraction among elements (mov-11 and mov-12, SM).
The nondimensional period of the oscillation, about $6$, and the
relaxation time, $0.1$, correspond to 15 minutes and 15 s,
respectively. They are close to the observed period of, respectively,
the cluster oscillation and the droplet oscillation [Fig.~\ref{small}(c)
and (d)].

On the other hand, if the forces are
homogeneous, that is, all the elements interacts and are propelled
similarly, they start to move in a circular motion as shown in
Fig.~\ref{simulation}(c) and (d) (mov-13, SM), and do not
show back-and-forth oscillation. Thus the inhomogeneity of the forces
plays a crucial role in the dynamic oscillatory state.

We could also reproduce other cluster
motions including a four droplets translation [cluster 1 in
Fig.~\ref{small}(a) and mov-14, SM] and a rotation of S-shaped chain
observed in other experiments (mov-15 and mov-16, SM).  However, so far
the wiggling motion seen in a larger cluster like the one shown in
Fig.~\ref{worm} has not been reproduced yet. In our model, the
interaction $k_{ij}$ and the propulsion $f_i$ are fixed throughout a
simulation. In the wiggling motion, however, the value of these
parameters might also change with time according to the situation.

 In conclusions, we found a self-propelled droplet system driven by the
solutal Marangoni effect, where the droplets self-organized into a
characteristic dynamic oscillatory state. The interaction among droplets
was essential for this dynamic oscillatory state, since this state
appeared only with neighboring particles and in a stage where a
single particle motion decayed mostly.

We tested a model which did not assume the homogeneous forces acting on
elements. The model reproduced a 5-element oscillation well, as well as
some behaviors of small clusters. This suggests the inhomogeneity of the
interaction as well as the propulsion forces, due to the Marangoni
effect mediated by the field around the elements, which does not
necessarily obey the superposition principle. This is essentially
different from a passive material where the interaction potential is
pre-determined.

Our system needs to be investigated further, especially in terms of the
active, inhomogeneous interaction, not only for understanding the
mechanisms of the dynamic ordering phenomena, but
also for designing life-mimicking systems. We believe that our
experimental system can be an ideal model system for this direction of
investigation, for its simplicity, robustness, and long-lasting motions.

\end{document}